\documentstyle[12pt,aasms]{article}
  \begin{document}
  \title {The H~I Environment of Nearby Lyman-alpha Absorbers}

  \author{J. H.~van Gorkom}
  \affil {Department of Astronomy, Columbia University, NY 10027}

  \author {C. L.~Carilli}  
  \affil {National Radio Astronomy Observatory, P.O. Box O, Socorro, NM 87801}

  \author{John T.~Stocke, Eric S.~Perlman\altaffilmark{1}, and J. Michael Shull
  \altaffilmark{2}}
  \affil {Center for Astrophysics and Space Astronomy,Department of 
  Astrophysical, Planetary, and Atmospheric Sciences, University of Colorado, 
  Boulder, CO 80309}

  \altaffiltext{1}{ now at Laboratory for High Energy Astrophysics, 
  NASA-Goddard Space Flight Center, Mail Code 660.2, Greenbelt, MD 20771}

  \altaffiltext{2}{also at JILA, University of Colorado and National Institute 
  of Standards and Technology} 

  \begin{abstract}
  We present the results of a VLA and WSRT search for H~I emission from the
  vicinity of seven nearby clouds, which were observed in Ly$\alpha$ absorption
  with HST toward Mrk~335, Mrk~501 and PKS~2155-304.  Around the absorbers, we
  searched a volume of $40' \times 40' \times$ 1000 km~s$^{-1}$; for one of the
  absorbers we probed a velocity range of only 600 km~s$^{-1}$. The H~I mass
  sensitivity (5 $\sigma$) very close to the lines of sight varies from $5 \times
  10^6$ M$_\odot$ at best to $5 \times 10^8$ M$_\odot$ at worst. 

  We detected H~I
  emission in the vicinity of four out of seven absorbers. The closest galaxy we
  find to the absorbers is a small dwarf galaxy at a projected distance of $68
  h^{-1}$ kpc from the sight line toward Mrk~335. This optically uncataloged
  galaxy has the same velocity ($V = 1970$ km~s$^{-1}$) as one of the absorbers,
  is fainter than the SMC, and has an H~I mass of only $4 \times 10^7$ M$_\odot$.
  We found a somewhat more luminous galaxy at exactly the velocity ($V = 5100$
  km~s$^{-1}$) of one of the absorbers toward PKS~2155-304 at a projected
  distance of $230 h^{-1}$ kpc from the sight line. Two other, stronger absorbers
  toward PKS~2155-304 at $V \approx 17,000$ km~s$^{-1}$ appear to be associated
  with a loose group of three bright spiral galaxies, at projected distances of
  300 to $600 h^{-1}$ kpc. These results support the conclusions emerging from
  optical searches that most nearby Ly$\alpha$ forest clouds trace the
  large-scale structures outlined by the optically luminous galaxies, although
  this is still based on small-number statistics. We do not find any evidence
  from the H~I distribution or kinematics that there is a physical association
  between an absorber and its closest galaxy. While the absorbing clouds are at
  the systemic velocity of the galaxies, the H~I extent of the galaxies is fairly
  typical, and at least an order of magnitude smaller than the projected distance
  to the sight line at which the absorbers are seen. On the other hand, we also 
  do not find evidence against such a connection. 

  In total, we detected H~I emission from five galaxies, of which two were
  previously uncataloged and one did not have a known redshift. No H~I emission
  was detected from the vicinity of the two absorbers, which are located in a
  void and a region of very low galaxy density; but the limits are somewhat less
  stringent than for the other sight lines. These results are similar to what has
  been found in optically unbiased H~I surveys. Thus, the presence of Ly$\alpha$
  absorbers does not significantly alter the H~I detection rate in their
  environment. 
  \end{abstract}

  \section{Introduction}

  The plethora of low column density, intervening Ly$\alpha$ absorption  lines in
  the spectra of high redshift QSO's (the ``Ly$\alpha$ forest'') was first
  recognized by Lynds (1971) and has been described in detail by Sargent 
  et al. (1980) and
  in many recent reviews and articles (Bajtlik 1993; Weymann 1993; Bechtold 1993;
  Rauch et al. 1992; Rauch et al. 1993; Lu, Wolfe, \& Turnshek 1991; Smette et
  al. 1992). A considerable amount has been learned about the nature of these
  systems at redshifts $z \ge 1.6$, the redshift above which Ly$\alpha$ is
  observable from the ground. These redshifts are generally too large to directly
  detect the absorbing objects in emission. However, more recently, a modest
  number of Ly$\alpha$ absorbers at low redshift have been detected in the UV
  (Morris et al. 1991;  Bahcall et al. 1991).  The proximity of these systems
  make them ideal targets for searches for H~I and optical emission from the
  vicinity of these clouds. One can also seek possible identifications of parent
  objects with which the clouds might be associated. 

  The first low-redshift Ly$\alpha$ forest absorption lines were discovered in
  the IUE spectrum of PKS~2155-304 by Maraschi et al. (1988). Since then, the HST
  has revealed many low column density absorbers in the nearby universe (Bahcall
  et al. 1991; Bahcall et al. 1993; Morris et al. 1991; Bruhweiler et al. 1993;
  Stocke et al. 1995; Shull et al. 1996). Statistical studies have been made to
  determine the relative correlation between these Ly$\alpha$ absorbing clouds
  and optical galaxies (Morris et al. 1991; Salzer 1992; Stocke et al. 1995;
  Lanzetta et al. 1995; Mo \& Morris 1994; Morris et al. 1993; Morris \& van den
  Bergh 1994). The current consensus is that the systems do correlate weakly with
  bright galaxies, but less so than these galaxies with other galaxies (Stocke et
  al. 1995). Lanzetta et al. (1995) find good evidence that some of the stronger
  Ly$\alpha$ absorbers are physically close to galaxies, but there are also
  examples of clouds with no optical galaxy to within a few Mpc (Stocke et al.
  1995; Shull et al. 1996). Morris et al. (1991) find, in one instance, an
  anti-correlation between a region of very high galaxy density and Ly$\alpha$
  forest absorbing clouds. Stocke et al. (1995) find more generally that: ``the
  higher equivalent width absorbers are distributed more like galaxies than the
  lower equivalent width absorbers, which are distributed in a manner
  statistically indistinguishable from clouds randomly placed with respect to
  galaxies.'' 

  In this paper, we present a search for H~I emission from the vicinity of seven
  nearby Ly$\alpha$ absorbers.  H~I imaging surveys routinely find gas-rich,
  optically uncataloged galaxies, and as such our search is complementary to the
  optical surveys mentioned above. In addition to searching for a possible parent
  population of the Ly$\alpha$ absorbers, the H~I morphology of galaxies close to
  the line of sight might betray hints of unusually large gaseous extents. In the
  first study of this kind (van Gorkom et al. 1993) a deep search for H~I
  emission was made around two Ly$\alpha$ clouds on the 3C~273 line of sight,
  which are located in the outskirts of the Virgo Cluster. No obvious
  associations between these two Ly$\alpha$ clouds and H~I emitting galaxies were
  found. The seven systems studied here are located in a wide range of cosmic
  environments. The absorbers seen along the sight line toward Mrk~501 are
  located in a void and a very low density region, respectively,  
  while the other five absorbers
  are located in regions of moderately high galaxy density, along the sight lines
  toward PKS~2155-304 and Mrk~335. The results of the H~I search near the sight
  line toward  Mrk~501 have already been presented by Stocke et al. (1995). Here,
  those observations will be presented in somewhat more detail. We describe the
  systems and observations in \S~2 and \S~3. We present the results in \S~4, 
  and in \S~5 we  briefly discuss their implications.

  Throughout this paper we adopt heliocentric velocities, using the optical
  definition, $V_{opt} = cz$, where c is the speed of light and the redshift is
  defined as $z= {\lambda-\lambda_0\over\lambda_0}$, where $\lambda$ and
  $\lambda_0$ are the observed and rest wavelengths, respectively.

  \section {The Systems} 

  The seven absorption-line systems have been discovered by different authors
  and their properties can be found in the literature. In Table 1 we summarize
  the sources against which they have been found, the heliocentric velocities
  of the lines, and the measured equivalent widths. In this Table we also
  give the projected distance to our H~I detection closest to the line of sight
  and the difference between the velocity of the absorption line and the 
  systemic velocity of the galaxy as derived from the H~I. 

  Two of the systems are seen in absorption against the BL Lac object Mrk~501,
  which is located in the ``Great Wall'' of galaxies at heliocentric velocity
  10,300 km~s$^{-1}$. The two absorbing systems are at 7530 km ~s$^{-1}$ and 4660
  km~s$^{-1}$ (Stocke et al. 1995). The 7530 km~s$^{-1}$ system is located in the
  void between us and the great wall, it has no cataloged optical galaxies within
  $4.5 h^{-1}$ Mpc, where $h$ is the Hubble constant in units of 100
  km~s$^{-1}$~Mpc$^{-1}$. The 4670 km~s$^{-1}$ system has a sparse chain  of
  galaxies located  to the southwest, the closest of which is $25'$ ($340 h^{-1}$
  kpc) off the line of sight to Mrk~501. 

  The detections of five, and possibly six, local Ly$\alpha$ forest lines have
  been reported for the sight line toward the BL Lac object PKS~2155-304
  (Bruhweiler et al. 1993; Maraschi et al. 1988; Allen et al. 1993). We did an
  H~I search around the three stronger lines, at 5100 km~s$^{-1}$, 16,488
  km~s$^{-1}$ and 17,088  km~s$^{-1}$ respectively. The last two velocities are
  uncertain by about 140 km~s$^{-1}$. The lower velocity system is on the edge of
  the Perseus -- Pisces supercluster; its nearest cataloged galaxy is
  ESO~466-G032, 15$'$ ($230 h^{-1}$ kpc) east of the sight line toward
  PKS~2155-304 with a systemic velocity of 5187  km~s$^{-1}$. The higher velocity
  systems have a number of cataloged galaxies at similar velocities not far from
  the line of sight. The closest is 2155-3033 (from the CfA redshift catalog), an
  Sb spiral 6.4$'$ ($305 h^{-1}$ kpc) to the SW of the line of sight to
  PKS~2155-304 and with a systemic velocity of 17,300 km~s$^{-1}$. 

  The third sight line that we investigated is toward Mrk~335, which has four 
  Ly$\alpha$  absorption lines (Stocke et al. 1995). Here, we observed the two
  lower velocity systems, which are close together in velocity at 1970 and 2290
  km~s$^{-1}$. These systems are located in a supercluster region. The closest
  cataloged galaxy is NGC~7817, an SAbc spiral, 46.5$'$ ($311 h^{-1}$ kpc) to the
  NE of the sight line toward Mrk~335 and with a systemic velocity of 2309
  km~s$^{-1}$. 

  The H~I column densities for the absorbing systems are all in the range of
  10$^{13-14}$ cm$^{-2}$ (assuming a Doppler parameter $b = 30$ km~s$^{-1}$),
  except for the complex absorption lines around 17,000 km~s$^{-1}$.  For these,
  a large column density, $\sim10^{18}$ cm$^{-2}$, has been derived (Maraschi et
  al. 1988), placing it in the lower column density  range of heavy element
  quasar absorption line systems. Thus far, no firm detections of metal lines
  have been made in this system, implying a metallicity $<$ 0.1 solar (Bruhweiler
  et al. 1993).

  \section {Observations and Data Processing}

  One of the sight lines, the one toward Mrk~501, was observed with the 
  Westerbork Synthesis Radio Telescope (WSRT), while the other searches
  were done with the Very Large Array (VLA). We describe the observations 
  and data processing for each of the sight lines separately. Table 2 summarizes
  the observations.

  \subsection{Mrk~501}

  The two Ly$\alpha$ absorption systems towards Mrk~501 were observed with the
  WSRT. Two 12 hour synthesis observations were made centered at 4885 km~s$^{-1}$
  with short spacings of 36 m and 54 m, and a single 12 hour synthesis
  observation, centered at 7740 km~s$^{-1}$. The total velocity range covered in
  each case was 1000 km~s$^{-1}$ at a resolution of 17 km~s$^{-1}$. Standard
  calibration of the data was performed using the NEWSTAR data reduction software
  package at NFRA. Further data editing, imaging, and analysis were performed
  using the AIPS package. 

  Mrk~501 is a radio continuum source, with an observed flux density at 1.4 GHz
  of 1.3 Jy. The source is unresolved at 15$''$ resolution, with a  position
  of 16$^{\rm h}$ 52$^{\rm m}$ 11.65$^{\rm s}$, 39$^{\circ}$ 50$'$ 27.1$''$ 
  (B1950). For
  the high-velocity system, the pointing center corresponded to the position of
  the continuum source, while the observations at 4675 km~s$^{-1}$ were made
  pointing 5$'$ to the southwest of Mrk~501 in order to increase sensitivity at
  the sparse chain of galaxies to the southwest. 

  Subtraction of the continuum emission from the line data was performed in two
  ways. The first involved linear fits in frequency to the calibrated complex
  visibilities (Cornwell et al. 1992). The second involved linear fits in
  frequency to the multi-channel image cube. The results from the two methods
  were similar, although the residual artifacts towards the edges of the field
  were worse for the image-plane subtraction (as expected), and the analysis
  presented herein relies on the uv-data continuum subtraction method. 

  Images of Mrk~501 were made  using ``natural'' weighting of the visibilities.
  The FWHM of the WSRT synthesized beam was $25'' \times 15''$. The image cubes
  were smoothed in velocity to 34 km~s$^{-1}$ resolution and visually inspected
  for 21~cm emission signal. No H~I signal is seen anywhere in the cube above
  five times the RMS in a given channel.   The image cube was then smoothed in
  both velocity and spatially to 68 km~s$^{-1}$ channel$^{-1}$ and to 38$''$
  spatial resolution. Again, no emission is seen anywhere in the cubes above five
  times the RMS at any velocity or spatial resolution. The noise values at
  various resolutions are listed in Table 3.

  \subsection{PKS~2155-304}

  PKS~2155-304 was observed twice with the VLA. The first set of observations
  was done with the 1 km array with an extended (3 km) north arm (DnC 
  configuration) to compensate for the southern declination of the source.
  The low-velocity system was observed for 4 hours, centered at 5100 km~s$^{-1}$,
  covering 560 km~s$^{-1}$ with a velocity resolution of 21 km~s$^{-1}$. The 
  high velocity systems were observed for a total of 10 hours. These observations
  were centered at 17,400 km~s$^{-1}$ in an effort to also include a possible
  Ly$\alpha$ system at 17,700 km~s$^{-1}$. However, more recent HST data,
  obtained with the Faint Object Spectrograph and G130H grating do not confirm
  the reality of that system (Allen et al. 1993). The total velocity range
  covered was 620 km~s$^{-1}$ with a velocity resolution of 23 km~s$^{-1}$. All
  those data suffered from rather serious interference. The high velocity system
  was reobserved with the VLA in the 1 km (D) configuration to make up for the
  loss of data due to interference and to cover also the Ly$\alpha$ system at
  16,488 km~s$^{-1}$. The time on source was again 10 hours. The observations
  were centered at 16,880 km~s$^{-1}$ and covered 1280 km~s$^{-1}$ with a
  resolution of 46 km~s$^{-1}$. Unfortunately, these observations were made
  during daytime and the data suffered rather badly from solar interference. 

  PKS~2155-304 is a radio continuum source, with a variable flux density. We
  measured a flux density of 0.45 Jy at 1.4 GHz. All observations were centered
  at the radio position of the BL Lac object, 21$^{\rm h}$ 55$^{\rm m}$
  58.30$^{\rm s}$, $-30^{\circ}~27'$ 54.4$''$ (B1950), Standard calibration
  procedures were followed, giving special care to the bandpass calibration. A
  bandpass calibrator was observed once every 2 hours, and for each data point we
  used the bandpass solution closest in time. Initially, we subtracted the
  continuum by making a linear fit in frequency to the calibrated complex
  visibilities, using the inner 75$\%$ of the band. After the continuum
  subtraction, the data were clipped to remove solar and man-made interference.
  Images were made of each of the observing runs separately. After inspection of
  the cubes to locate the channels with H~I emission, the subtraction of the
  continuum was redone, making a fit to the line free channels only. 

  For the low velocity system, images were made using natural weighting, resulting
  in a FWHM of the synthesized beam of $66'' \times 39''$. For the high-velocity
  systems the data of the various observing runs were combined in the UV plane,
  after Hanning smoothing the data of the first run down to the resolution of the
  second run. Images were made using natural weighting, resulting in a
  synthesized beam of $97.8'' \times 47.2''$. Several galaxies were detected in
  H~I, one in the low-velocity data and three in the high-velocity data. These
  results are described in \S~4.

  \subsection{Mrk~335}

  We observed Mrk~335 with the VLA in the 1 km (D) array for 4 hours in total.
  The observations were centered at 2135 km~s$^{-1}$ in between the velocities of
  the two Ly$\alpha$ absorbers, covering a range of 1000 km~s$^{-1}$ with a
  velocity resolution of 25 km~s$^{-1}$. The observations were centered at the
  optical position of the Seyfert galaxy, at 00$^{\rm h}$ 03$^{\rm m}$ 45.2$^{\rm
  s}$, 19$^{\circ}$ 55$'$ 28.6$''$ (B1950). The Seyfert galaxy itself is not a
  radio continuum source. We subtracted background continuum sources by making a
  linear fit to the calibrated complex visibilities. Images were made using
  natural weighting, resulting in a synthesized beam of $62. 8'' \times 52.7''$.
  One previously uncataloged galaxy was discovered in H~I. We describe this
  result in the next section. 

  For the H~I detections, we made images of the total hydrogen emission by
  smoothing the data spatially and in velocity, using the smoothed cube as a mask
  for the full resolution data. Only pixels above 2 $\sigma$ in the smoothed cube
  were used in the sum. Throughout this paper we use images of the digitized POSS
  to show overlays of neutral hydrogen emission on optical images.

  \section{Results}

  \subsection {Upper Limits}

  Although the more interesting result of this work is the detection of so many
  galaxies, the upper limits we can place on any H~I in emission close to the
  Ly$\alpha$ absorbers are important too. In Table 3 we list the mass and column
  density limits for each of the systems. These are 5 $\sigma$ limits at the
  field center. Farther from the center, they need to be corrected for the change
  in primary beam response. The primary beam pattern is roughly Gaussian with a
  FWHM of 36$^{\prime}$ and 30$^{\prime}$ for the WSRT and the VLA respectively.
  For example, the VLA limits are a factor of two worse at 15$^{\prime}$  from
  the field center, and 10 times worse at 25$'$. At the position of the sparse
  group to the SW of Mrk~501, the mass and column density limits are a factor of
  five  worse than at the field center. Although these limits put serious
  constraints on the presence of small, gas-rich dwarf galaxies very close to
  each of the lines of sight, the small dwarf detected toward Mrk 335 at 12$'$
  from the line of sight could only have been detected in the observations toward
  Mrk~335 and in the low-velocity system toward PKS~2155-304. Interestingly, in
  both these observations (and only those) a small galaxy was found very close to
  the velocity of the absorber, but at rather large projected distance from the
  line of sight. It should also be noted that in the sight line toward Mrk~501,
  the one observation in which no galaxies were detected at all the surface
  brightness sensitivity is ten times worse than that of the other observations.
  In fact, some of the most gas-rich, but low H~I surface density galaxies such
  as Malin 1 could have escaped detection in those data.

  \subsection{ H~I Detections}

  \subsubsection {Mrk~335}

  Perhaps the most exciting result of these observations is the detection of the
  small dwarf near the sight line toward Mrk~335. Contour images of the velocity
  channels are shown in Fig. 1. Although the emission is not resolved in
  individual channels, the peak of the emission clearly shifts in position with
  velocity. The maximum displacement of the peaks gives us a lower limit to the
  H~I extent, which is $45''$ or $4.2 h^{-1}$ kpc. A position -- velocity profile
  along the major axis of the galaxy (at a PA of $0^{\circ}$) is shown in Fig. 2.
  To put things in perspective, we show in Fig. 3 an overlay of the total H~I
  emission onto an optical image, which includes Mrk~335 as well. Although the
  H~I is more extended than the optical emission of this tiny dwarf galaxy, there
  is a huge distance ($68 h^{-1}$ kpc) between the lowest H~I contour (at $4.8
  \times 10^{19}$ cm$^{-2}$) and the sight line toward Mrk~335. The H~I
  properties derived from this observation are summarized in Table 4. The
  systemic velocity as derived from the H~I differs by only 20 km s$^{-1}$ from
  that of the higher column density absorber at 1970 km s$^{-1}$.

  \subsubsection {PKS~2155-304}

  At low velocities, we detect H~I emission from ESO~466-G032, located $230
  h^{-1}$ kpc to the east of the line of sight to PKS~2155-304. This is the
  closest optically cataloged galaxy to the line of sight. The H~I is seen over
  the velocity range 5080 km~s$^{-1}$ to 5130 km~s$^{-1}$. Contour images of the
  velocity channels at 11.7 km~s$^{-1}$ resolution are shown in Fig. 4. Crosses
  mark the optical position of the H~I emitting galaxy. The systemic velocity as
  derived from the H~I profile is 5100 km~s$^{-1}$, significantly less than the
  reported optical value of 5187 km~s$^{-1}$. Optically, the galaxy looks rather
  disturbed, with a spiral arm (or tidal feature?) extending to the southwest.
  The H~I emission is just barely above the noise. Although it appears to be
  slightly extended in the direction of the extended optical feature, the
  synthesized beam is extended in the same direction. There is a hint of rotation
  along the optical major axis of the galaxy, with the receding side to the
  northeast (Fig. 4). The angular resolution of the current data is not
  sufficient to say any more about the H~I morphology; the signal is only barely
  resolved. The H~I parameters are summarized in Table 4. Again, to put things in
  perspective we show in Fig. 5 an H~I overlay on an optical image including both
  the galaxy and the BL Lac object. 

  The high-velocity data cube covers the velocity range from 16,240 km s$^{-1}$
  to 17,720 km s$^{-1}$, but the quality varies across the band. At velocities in
  excess of 17,500 km s$^{-1}$, the data quality is poor. We detect H~I emission
  from three galaxies. The galaxy closest to the line of sight, 2155-3033, is at
  a projected distance of $305 h^{-1}$ kpc to the south. The H~I emission from
  this galaxy is barely above the noise. A weak signal is seen over velocities
  from 16,900 km~s$^{-1}$ to 17,300 km~s$^{-1}$, with no detectable H~I in the
  two middle channels (Fig. 6). The systemic velocity derived from the H~I is
  17,100 km~s$^{-1}$, the optical redshift gives a velocity of 17,300
  km~s$^{-1}$. 

  A much stronger H~I emitter is found to the southeast at a projected distance
  of $610 h^{-1}$ kpc. This is an IRAS source, F21569-330, with a hitherto
  unknown redshift. The H~I emission can be seen over a velocity range from
  16,650 km~s$^{-1}$ to 16,925 km$^{-1}$ (Fig. 7). The kinematic major axis lies
  along the optical major axis at a PA of $-40^{\circ}$, with the north side
  receding. A position velocity profile along the major axis is shown in Fig. 8. 

  Almost due north of PKS~2155-304 we detect an optically uncataloged galaxy at a
  projected distance of $515 h^{-1}$ kpc from the line of sight. H~I can be seen
  in emission from 17,063 km s$^{-1}$ to 17,248 km s$^{-1}$ (Fig. 9). The
  kinematic major axis is east -- west and a position velocity profile along this
  axis is shown in Fig. 10. Optically the galaxy looks distorted, slightly
  extended to the northwest with what are possibly two dwarf companions. The H~I
  parameters derived from these observations are listed in Table 4. Finally, to
  put things in perspective, we show in Fig. 11 an overlay of the H~I emission of
  the three galaxies detected in the high velocity cube on an optical image,
  which shows, in addition to the galaxies,  PKS~2155-304.

  \section {Discussion}

  Our search for H~I emission from the vicinity of nearby Ly$\alpha$ absorbers
  has resulted in the detection of five galaxies, two of which were previously 
  uncataloged. We shall first discuss whether this detection rate is unusual
  in any sense: does the presence of a Ly$\alpha$ absorber increase the chance of
  finding H~I in emission? Following that, we shall discuss what we have learned
  from the detections, and whether there is any indication that the H~I in
  emission is related to the Ly$\alpha$ absorption. Finally we will discuss
  whether the present observations have illuminated the nature of the nearby
  Ly$\alpha$ absorbers. 

  A variety of data is available to assess whether our detection rate of H~I-rich
  systems is unusual in any way. Briggs (1990) summarized the results of all
  major unbiased H~I surveys, while the currently most accurate H~I luminosity
  function has been constructed by Rao \& Briggs (1993). More specific and
  directly comparable to our result is the work by Weinberg et al. (1991) and
  Szomoru et al. (1994, 1996), who used the VLA to do unbiased H~I surveys in
  environments of differing galaxian density. Szomoru et al. (1994) probed voids
  as well as supercluster environments and compared fields centered on optically
  known galaxies and optically blank fields. 

  For simplicity sake, we take as the volume searched in our survey the region
  within a radius of 22$'$ from the pointing center. This is the 20$\%$ point of
  the primary beam; beyond that, the shape of the beam is highly uncertain. Our
  search within that volume is complete to the 5 $\sigma$ limits listed in Table
  3, multiplied by five (correcting for the primary beam response). The volume
  searched down to a mass limit of $2 \times 10^9 M_{\odot}$ of H~I is $60
  h^{-3}$ Mpc$^3$. In this volume, we detected three galaxies with H~I masses of
  a few times 10$^9$ M$_{\odot}$ of H~I, giving a density of 0.05 $\pm$ 0.02
  Mpc$^{-3}$. The volume searched down to $5 \times 10^8 M_{\odot}$ of H~I is
  much smaller, $6.8 h^{-3}$ Mpc$^3$. Two galaxies of even smaller masses were
  detected in that volume, bringing the galaxy density in the mass range of
  10$^7$ M$_{\odot}$ to $ 5 \times 10^8$ to $0.3 \pm 0.2 h^3$ Mpc$^{-3}$. These
  H~I mass densities are quite consistent with the results of Briggs (1990), who
  found a density of $0.07 h^3$ Mpc$^{-3}$ and $0.1 h^3$ Mpc$^{-3}$ for H~I
  masses of a few times 10$^9$ M$_{\odot}$ and 10$^8$ M$_{\odot}$ respectively.
  Only a small fraction of the volume searched, $7 h^{-3}$ Mpc$^3$, is in a true
  void; the remaining volume is more characteristic of supercluster densities.
  Weinberg et al. (1991) found a cumulative space density of $0.13 h^3$
  Mpc$^{-3}$ for gas rich dwarfs above 10$^8$ M$_{\odot}$ of H~I in the Perseus
  -- Pisces supercluster, a result similar to ours. In conclusion, although the
  statistics are small, it appears that the presence of nearby Ly$\alpha$ forest
  absorbers has not significantly altered the detection rate of H~I emitting
  objects in either the void or the high-density regions.

  \subsection{The Markarian 335 Sight Line }

  Nevertheless, the detections of two galaxies almost exactly at the velocity of
  their nearby  Ly$\alpha$ absorbers, and the fact that  these are the only
  detections  in a 500 and 1000 km s$^{-1}$ range for PKS~2155-304 and Mrk~335
  respectively, beg the question whether there is a possible association between
  the galaxies and the absorbers. The most tantalizing system is the uncataloged
  dwarf galaxy at a projected distance of $68 h^{-1}$ kpc from the line of 
  sight toward Mrk~335. The absolute magnitude of this dwarf has been estimated
  to be $M_R  \approx -15.4$ based upon a linear extrapolation of the calibration
  supplied by eight, nearby HST guide stars on the digitized version of the 
  POSS-E plate material. This approximate magnitude assumes that 
  H$_0$ = 100 km s$^{-1}$ Mpc$^{-1}$ despite the location of this object within 
  the bounds of the local supercluster. 
  The proximity, spatially and in velocity, of the Ly$\alpha$ absorber
  and dwarf irregular could mean no more than that they originated in a common
  larger scale structure, e.g., a filament (Cen et al. 1994; 
  Hernquist et al. 1996; Miralda-Escud\'e et al. 1996; M\"ucket et al. 1996). 
  Alternatively, it could imply
  that the Ly$\alpha$ cloud is gravitationally bound to the dwarf. Perhaps, it is
  still falling in or has been ejected in a galactic wind. The kinematic
  structure does not help to choose between these various possibilities.  The
  minimum dynamical mass in the Ly$\alpha$ cloud + dwarf galaxy system needed to
  make it a bound system can be derived by requiring that the total kinetic
  energy of the system is less or equal to the potential energy. Thus, 

  $$ M_{dyn} = {1\over 2G} (\Delta V)^2 \Delta R \eqno(1) $$
  where G is the gravitational constant, $\Delta V$ the line-of-sight velocity
  difference and $\Delta R$ the projected distance between absorber and galaxy.
  The minimum dynamical mass is 4.6 $\times$ 10$^9$ M$_{\odot}$. A crude estimate
  of the mass of the dwarf galaxy can be obtained from the H~I kinematics.
  Assuming a total H~I extent of 3.8 kpc and a rotation velocity of 55 km
  s$^{-1}$, we find $M = 6.7 \times 10^8~M_{\odot}$. Thus, in order for the
  Ly$\alpha$ cloud to be bound to the galaxy, the dwarf needs to be embedded in a
  massive dark halo. 

  An intriguing possibility is that the Ly$\alpha$ absorption arises from within
  a mostly ionized gas disk of the dwarf galaxy. Maloney (1992) and Stocke et al.
  (1995) discussed the possibility that nearby Ly$\alpha$ lines could be produced
  by gas at large radii in the disks of spiral and irregular galaxies. Maloney
  concludes that the unexpectedly large number of low-redshift Ly$\alpha$
  absorption lines seen with HST toward 3C~273 can be produced by extended
  ionized gas disks in the halos of spiral and irregular galaxies. The observed
  frequency of absorbers requires that either L$_*$ galaxies have huge (several
  hundred kpc) halos or, more plausibly, that the decrease of absorption cross
  section with declining luminosity is slow enough for low-luminosity galaxies to
  dominate the integrated cross section.  Shull et al. (1996) make the case
  that, for absorbers of mean radius $R = (100~{\rm kpc})R_{100}$,
  only dwarf galaxies have the comoving space densities,
  $$ \phi_0 \approx (0.9~{\rm Mpc}^{-3}) R_{100}^{-2} h \eqno(2) $$
  necessary to explain the frequency of low-redshift Ly$\alpha$ clouds. For
  reference, this density is over 20 times that of $L_*$ galaxies, recently
  estimated as $\phi(L_*) \approx 0.04 h^3$ Mpc$^{-3}$ (Marzke et al. 1994).

  A first hint that low-luminosity galaxies may indeed dominate the cross section
  comes from the discovery (Barcons et al. 1995) of possibly corotating
  Ly$\alpha$ absorption at large (50 kpc) projected distance from two small,
  late-type spiral galaxies. Contrary to the cases found by Barcons et al.
  (1995), we have no evidence for corotation of the ionized gas with the galaxy,
  the Ly$\alpha$ absorption occurs too close to the minor axis of the galaxy.
  The small velocity difference between the absorber and the systemic velocity of
  the galaxy is of course not inconsistent with corotation, but at that large
  distance anything that is bound to the galaxy is expected to have a velocity
  close to the systemic velocity of the galaxy. 

  Although Stocke et al. (1995) discussed the problems with huge ionized halos
  around a single bright spiral galaxy, it might be possible that the
  Ly$\alpha$ absorption at $68h^{-1}$ kpc from the dwarf arises in an ionized
  halo. Using the model calculations of Maloney (1992) and Dove \& Shull (1994),
  we find that an ionized 80--100 kpc halo is not at all unlikely for a dwarf
  galaxy. For example, consider a spherical dark-matter halo, with central
  density $\rho_0$, core radius $r_c$, and asymptotic halo velocity $v_A = (4 \pi
  G \rho_0 r_c^2)^{1/2} \equiv (50~{\rm km~s}^{-1})v_{50}$. The 
  three-dimensional velocity dispersion is $\langle v^2 \rangle = (3/2)v_A^2$. In
  equilibrium, the gaseous hydrogen will settle into an atmosphere above the disk
  plane, with density 

  $$ n_H(z) = n_H(0) \exp \left[ {-z^2}\over{2 \sigma_h^2} \right] \;,\eqno(3)$$
  where $\sigma_h \approx R(\sigma_z/v_A)$ is the vertical scaleheight and
  $\sigma_z = (18.1~{\rm km~s}^{-1}) T_{4.3}^{1/2}$ is the thermal velocity of
  hydrogen at temperature $T = (10^{4.3}~{\rm K}) T_{4.3}$.  At a
  radius $R = (100~{\rm kpc}) R_{100}$ from the galactic center, we may
  express the total hydrogen density at the disk midplane as,

  $$   n_H(0) = { {N_H v_A} \over {(2 \pi)^{1/2} \sigma_z R}}
     = (3.57 \times 10^{-6}~{\rm cm}^{-3}) N_{18} v_{50} R_{100}^{-1}
      T_{4.3}^{-1/2}  \; ,\eqno(4)
  $$
  where $N_H = (10^{18}~{\rm cm}^{-2}) N_{18}$ is the total column density of 
  hydrogen, neutral and ionized, integrated through the disk. 

  In the optically-thin limit, the neutral hydrogen density is set by
  photoionization equilibrium, 
  $$
     n({\rm H}^0) = {{ n_e n_{H^+} \alpha_H^{(1)} } \over {\Gamma_H} } \; ,
  \eqno(5)$$
  where we adopt a case-A radiative recombination rate coefficient
  $$
     \alpha_H^{(1)} = (2.48 \times 10^{-13}~{\rm cm}^3~{\rm s}^{-1}) 
       T_{4.3}^{-0.726}  \eqno(6) 
  $$
  and a hydrogen photoionization rate $\Gamma_H = (2.64 \times
  10^{-14}~{\rm s}^{-1}) I_{-23}$. Here, we express the metagalactic ionizing
  radiation field, $I_{\nu} = I_0 (\nu/\nu_0)^{-\alpha_s}$, with 
  $$I_0 = (10^{-23}~{\rm ergs~cm}^{-2}~{\rm s}^{-1}~{\rm Hz}^{-1} 
     {\rm sr}^{-1}) I_{-23} \eqno(7) 
  $$
  at $h \nu_0 = 13.6$ eV and adopt a spectral index 
  $\alpha_s \approx 1.5$.  For a fully ionized gas
  with $n_{He}/n_H = 0.1$ and $n_e/n_H = 1.2$, we find $n({\rm H}^0) = (11.25)
  n_H^2 T_{4.3}^{-0.726} I_{-23}^{-1}$ and 
  $$
     N_{HI}(R) = (2.84 \times 10^{13}~{\rm cm}^{-2}) N_{18}^2 v_{50}
	  R_{100}^{-1} T_{4.3}^{-1.23} I_{-23}^{-1}   \; .\eqno(8)
  $$
  Note that $N_{HI}$ is proportional to $N_H^2/R$, owing to the $n_H^2$
  dependence of the recombinations that form H~I. In the inner portions of disks,
  where the gas layer is optically thick to external ionizing radiation, the H~I
  radial distribution, $N_{HI}(R)$, closely tracks that of total hydrogen,
  $N_H(R)$. However, in the extended disk, beyond the radius at which the
  integrated column of H~I drops below several times $10^{19}$ cm$^{-2}$, the
  disk becomes optically thin to ionizing radiation, and the above
  photoionization analysis applies.  In this regime, the radial H~I distribution
  falls off as $N_H^2/R$, which in an exponential gaseous disk results in a very
  sharp falloff in H~I. However, in disks with power-law gaseous distributions,
  $N_H(R) \propto R^{-\Gamma}$ ($1 \leq \Gamma \leq 2$), the H~I decreases with
  radius as $N_{HI}(R) \propto R^{-(2\Gamma+1)}$. 

  Let us now compare these expectations to the observed Ly$\alpha$ absorbers,
  which typically have columns N(H~I) $\approx 10^{13-14}$ cm$^{-2}$. From eq.
  (8), if the HST-observed Ly$\alpha$ cloud has H~I column density
  $10^{13.5}~{\rm cm}^{-2}$ at $R = (68~{\rm kpc})h^{-1}$, the gaseous disk at
  that radius must have a total hydrogen column density 
  $$
     N_H(R) = (0.87 \times 10^{18}~{\rm cm}^{-2}) v_{50}^{-1/2}
	h^{-1/2} I_{-23}^{1/2} T_{4.3}^{0.61} \; ,\eqno(9)
  $$
  assuming the optically-thin limit. The VLA observations of the dwarf H~I galaxy
  toward Mrk~335 yield an H~I column of $N_0 = 2.4 \times 10^{19}$ cm$^{-2}$ at
  $1'$, corresponding to $R_0 = 5.73h^{-1}$ kpc at the recessional velocity of
  1970 km~s$^{-1}$. If we assume that the radial distribution of total hydrogen
  column density is $N_H(R) = N_0 (r/R_0)^{-1}$ (i.e., $\Gamma = 1$) and
  integrate over radii $R_0 < r < R_{\rm max}$, where $R_{\rm max} = 68 h^{-1}$
  kpc, the total gaseous mass of the extended disk is 
  $$
     M_{\rm disk} = 2 \pi N_0 \mu R_0 (R_{\rm max} - R_0) = 
     (6 \times 10^{8}~M_{\odot}) h^{-2} \; ,\eqno(10)
  $$
  where we adopt a mean molecular mass $\mu = 1.4m_H$.  

  The mass of the isothermal dark-matter halo required to confine the gas
  in equilibrium is given by,
  $$
     M_{\rm halo} = { {2 \langle v^2 \rangle R}\over{G}} = (4 \times 10^{10}~
	 M_{\odot}) v_{50}^2 h^{-1}  \; .\eqno(11)
  $$
  The ratio of halo mass to gas mass is therefore $66 v_{50}^2 h$,
  not dissimilar from values in other galaxies.

  \subsection{The PKS~2155-304 Sight Line }

  A possible physical association between the Ly$\alpha$ absorber at 5100
  km~s$^{-1}$ seen toward PKS~2155-304 and ESO~466-G032 is even less clear cut.
  The projected distance between the two is $230 h^{-1}$ kpc. Lanzetta et al.
  (1995) find that only 1 out of 9 galaxies at projected distances larger than
  $160 h^{-1}$ kpc from the line of sight toward HST spectroscopic target QSO's
  gives rise to associated Ly$\alpha$ absorption, while 5 out of 5 galaxies at
  distances less than $70 h^{-1}$ kpc give rise to associated Ly$\alpha$
  absorption. However more recent data by Bowen et al. (1996) and Le Brun et al.
  (1996) show that the covering factor of galaxies between 50 and $300 h^{-1}$
  kpc is roughly 0.5 for equivalent widths larger than 0.3 \AA. Thus, our
  discovery does not seem that unusual. 

  The galaxy is a small Sb spiral with $M = -19.2$. It would be quite
  extraordinary if it had a gaseous halo extending out to 250 kpc or so. The long
  dynamic time scales at those distances make it unlikely that the gas has
  virialized or settled into a disk (Stocke et al. 1995). Note, however, that
  Zaritski \& White (1994) find that isolated spirals  have dark halos extending
  out to 200 or 300 kpc, thus even if the Ly$\alpha$ absorber is not part of a
  halo of ionized gas, it may still sit in the potential of the galaxy. As in the
  case of Mrk~335, the kinematics of the galaxy do not help to elucidate the
  situation, the Ly$\alpha$ absorber is at the systemic velocity of the galaxy. 

  In the high-velocity range toward PKS~2155-304, we see three galaxies at rather
  large projected distances from the line of sight. The three galaxies have a
  mean velocity of 17,021 km~s$^{-1}$ and a line of sight velocity dispersion of
  138 km~s$^{-1}$, typical of a loose group of galaxies. The Ly$\alpha$
  absorption at 17,100 km~s$^{-1}$ occurs close to the mean velocity. In this
  case, it seems more plausible that the absorption arises in some general
  intergalactic gas within a small group, rather than from one galaxy in
  particular (Mulchaey et al. 1993, 1996). 
  The galaxy to the southwest is in projection
  closest to the line of sight toward the quasar at a distance of $305 h^{-1}$
  kpc. Thus, if the absorption arises in the halo of the nearest galaxy, the halo
  would have to extend well beyond 300 kpc. The existence of such a huge ionized
  halo is problematic by itself (Stocke et al. 1995), in a group environment it
  will definitely not survive. Of course the intragroup gas may be just that, the
  remains of the shredded halos. It is interesting that the absorption at 17,100
  km~s$^{-1}$ is one of the stronger absorption systems. The situation is
  reminiscent of the absorbers found toward 3C~273 in the outskirts of the Virgo
  cluster (Bahcall et al. 1991; Morris et al. 1991). Several galaxies are found
  at distances of 200 to 300 kpc from the sight line, but it is not possible to
  associate the absorbers with individual galaxies (Morris et al. 1993; Salpeter
  and Hoffman 1995; Rauch et al. 1996). As in the present case, the absorbers 
  toward
  3C~273 have slightly higher column density. Although the current data for the
  PKS 2155-304 system, plagued as they are by interference, are not sensitive
  enough to detect small gas rich dwarfs, such as the one found toward Mrk 335,
  the 3C~273 data definitely rule out the existence of such dwarfs within 100 kpc
  from the line of sight toward 3C~273 (van Gorkom et al. 1993). 

  The absorption at 16,488 km~s$^{-1}$ seems more difficult to explain. It is
  quite far off from the mean velocity of the group. The H~I emission from 
  F21569-330 at a projected distance of $610 h^{-1}$ kpc comes closest in
  velocity going down to 16,650 km~s$^{-1}$, but not only is the distance
  to the sight line huge, the approaching side of the galaxy is on the far
  side  from the sight line to the quasar. Since for this group we have only
  sampled the upper end of the H~I mass function it is not unlikely that
  other galaxies are present which have H~I at the same velocity as the 
  absorber. Thus it seems plausible that this absorber is associated with
  the group as well.  

  In a recent paper Lanzetta, Webb and Barcons (1996) report the identification
  of a group of galaxies at a redshift of 0.26, that produce a complex of
  corresponding Ly $\alpha$ absorption lines. As in the present case one
  of the absorption systems has a slightly higher column density. Thus it 
  seems that Ly $\alpha$ absorption arising in intra group gas may be quite
  common and that the colomn densities in  those environments may be somewhat
  higher than what is seen near more isolated galaxies.

  \subsection{The Nature of Nearby Ly$\alpha$ Absorbers}

  A search for H~I emission from the vicinity of nearby Ly$\alpha$ absorbers has
  resulted in the discovery of some low luminosity galaxies at the same redshift
  of some of the absorbers, but at large (70 to 250 kpc) projected distances from
  the sight line. In addition a group of galaxies was found near two absorbers at
  a velocity of about 17000 km~s$^{-1}$. This confirms previous suggestions that
  nearby Ly$\alpha$ absorbers are weakly correlated with galaxies. Individual,
  more isolated, galaxies may have absorbers that are physically associated with
  them, either due to infall, galactic winds or tidal disturbances. In regions of
  higher galactic density this material may be stripped from individual galaxies
  and distributed more uniformly through the intergalactic medium. 

  This project was undertaken in the hope of finding a possible parent population
  associated with nearby Ly$\alpha$ absorbers: either luminous galaxies with very
  extended gaseous envelopes or very low luminosity, but gas rich galaxies, which
  could have escaped optical detection. We did indeed find a few optically
  uncataloged galaxies, which turned out to be very close in velocity to, but at
  large projected distances from the Ly$\alpha$ absorbers. Two of the absorbers
  appear associated with a group of galaxies. These results are very similar to
  searches for a possible parent population at optical wavelengths. Most
  absorbers at low redshift appear to coincide with the large scale structure
  outlined by the more luminous galaxies (Rauch et al. 1996; Stocke et al. 1995)
  as was first suggested by Oort (1981). 

  One of the outstanding questions is  whether the nearby Ly$\alpha$ absorbers
  are actually connected to individual galaxies or simply coincide in redshift.
  Our observations do to a certain extent strengthen the idea that at least some
  nearby Ly$\alpha$ absorbers may be arising in mostly ionized halos of 
  individual galaxies.
  The discovery in our most sensitive observation of a very small galaxy at $68
  h^{-1}$ kpc from the line of sight toward Mrk~335, suggests that: (1) deeper
  surveys may turn up more such candidates; and (2) the suggestion (Maloney
  1992; Shull et al. 1996) that the ionized halos of smaller galaxies may
  contribute significantly to the total Ly$\alpha$ absorbing cross section in the
  nearby universe may be valid. The strongest evidence for a physical connection
  between nearby Ly$\alpha$ absorbers and galaxies comes from the statistical
  work by Lanzetta et al. (1995), who showed that within $70 h^{-1}$ kpc from a
  galaxy there is a 100\% chance of detecting Ly$\alpha$ absorption. There are
  however also some Ly$\alpha$ clouds with no optical galaxy
  within a few Mpc (Stocke et al. 1995; Shull et al. 1996). In one case, an
  anticorrelation between a region of high galaxy density and Ly$\alpha$ forest
  absorbing clouds has been found (Morris et al. 1991). 
  The difference between these results may primarily be due to a difference 
  in cloud column densities. The Lanzetta et al (1995) absorbers have 
  significantly higher H~I column densities than those studied by Morris et 
  al. (1991), Stocke et al. (1995) and Shull et al. (1996). 

  Even if a physical
  connection is apparent, it is not obvious what the nature of the extended gas
  is. While Barcons et al. (1995) find in two cases Ly$\alpha$ absorption that
  could possibly be interpreted as arising in a corotating halo, there are
  several examples of even higher column density absorbers and metal lines, where
  the gas, although clearly associated with galaxies, is not corotating, but
  instead arises in tidally disturbed gas (e.g., Womble 1992; Carilli \& van
  Gorkom 1992; Bowen et al. 1995). Thus the final verdict is not out yet. Low
  column density gas is likely to be found near galaxies, and many possible
  scenarios can bring it there: retarded infall, outflow, corotating ionized
  disks, tidal disturbances. Quite likely, all of these occur. Perhaps most
  puzzling are the clouds that don't have any galaxies within many Mpcs. These
  may arise in the filamentary structures as produced in simulations of
  gravitational structure formation (Cen et al. 1994; Hernquist et al. 1996;
  Miralda-Escud\'e et al. 1996;
  M\"ucket et al. 1996), or they may be associated with as yet undetected dwarf
  galaxies. 

  This brings us to the final question, what is the connection, if any, between
  the high redshift Ly$\alpha$ forest clouds and nearby Ly$\alpha$ absorbers? As
  was pointed out by Rauch et al. (1996), the size estimate for coherent
  Ly$\alpha$ absorption at large redshifts (Bechtold et al. 1994; Dinshaw et al.
  1994; Fang et al. 1996) is typically larger than the transverse separation
  between galaxies near low redshift Ly$\alpha$ absorbers. This not only argues
  against single extended galactic disks or halos as the main origin for the
  coherent absorption on large scales seen at high redshifts. It also argues
  against the low redshift absorbers being physically the same as the high
  redshift absorbers. However, it may well be that at low redshifts galaxies and
  absorption systems trace the general matter distribution on large scales,
  a hypothesis that can only be tested statistically using a large ensemble
  of nearby Ly$\alpha$ clouds. 

  \acknowledgments

  JHvG thanks John Hibbard for comments on an earlier version of this manuscript
  and for stimulating discussions on the nature of Ly$\alpha$ clouds.
  We are grateful to Tony Foley for help with the WSRT observations. We thank the
  NFRA (Netherlands Foundation for Research in Astronomy) and the NRAO (National
  Radio Astronomy Observatory) for allocation of observing time. The NRAO is
  operated by Associated Universities, Inc. under a cooperative agreement with
  the National Science Foundation. This research was partly supported by NSF
  grant AST 90-23254 to Columbia University, a NOVA research fellowship to CLC at
  the University of Leiden, and through a HST Guest Observer grant
  (GO-3584.01-91A) and the NASA Astrophysical Theory program (NAGW-766) at the
  University of Colorado. JHvG thanks the Astronomy Dept of Caltech, where part
  of this work was done, for hospitality, Wal Sargent and Nick Scoville for 
  financial support, and Shri Kulkarni for entertainment. This research has made
  use of the NASA/IPAC extragalactic database (NED) which is operated by the
  JPL, Caltech, under contract with NASA.

\clearpage

\begin{figure}
\caption{ Contour images of the velocity channels for the sight line
toward Mrk~335. Only the area around the uncataloged dwarf, which was detected
in H~I emission is shown. 
The optical position of the galaxy center is shown
with a cross, heliocentric velocities in km s$^{-1}$
are indicated in the top right corner of each panel. The ellipse in the 
top left panel is the size of synthesiszed beam. Contour levels are 
-2, -1, 1, 2, 3, 4, 5, 6, 7 mJy per beam. Negative contours are dashed.}
\end{figure}

\begin{figure}
\caption{ A position velocity plot along the major axis, taken at a
position angle of 0$^{\circ}$, of the dwarf in the Mrk~335 field. Contour 
levels
are at -1.2, 1.2, 2.4, 3.6, 4.8, 6.0 mJy per beam.}
\end{figure}

\begin{figure}
\caption{ An overlay of the H~I emission (contours) on an image
of the digitized POSS showing Mrk~335, lower left corner and the dwarf
detected in H~I. The contour interval is 4.8 $\times$ 10$^{19}$ cm$^{-2}$.
The arrow points to Mrk~335.}
\end{figure}

\begin{figure}
\caption{ Contour images of the velocity channels of ESO~466-G032, which
is located close to the sight line of PKS~2155-304 at the same velocity as the
absorber at 5100 km~s$^{-1}$. The optical position of the galaxy center is
shown with a cross, heliocentric velocities in km s$^{-1}$ are indicated in the
top right corner of each panel. The ellipse in the top left panel is the size
of synthesized beam. Contour levels are -1.3, 1.3, 2.6, 3.9, 5.2 mJy per beam,
negative contours are dashed.} 
\end{figure}

\begin{figure}
\caption{ An overlay of the HI column density distribution (contours)
of ESO~466-G032 on an optical image (greyscale), which shows both the galaxy 
and  PKS~2155-304. The contour interval is 4.2 $\times$ 10$^{19}$
cm$^{-2}$. The arrow points to PKS~2155-304.}
\end{figure}

\begin{figure}
\caption{ Contour images of the velocity channels of the galaxy
2155-3033, south of PKS~2155-304. The optical position of the galaxy center
is shown with a cross, heliocentric velocities in km~s$^{-1}$ are indicated in
the top right corner of each panel. The ellipse indicates the size of the
synthesized beam. Contour levels are -1.2, -0.6, 0.6, 1.2, 1.8 mJy per beam.
Negative contours are dashed.}
\end{figure}

\begin{figure}
\caption{ Contour images of the velocity channels of F21569-330,
a galaxy to the south east of PKS~2155-304. The optical position of the galaxy
is shown with a cross, heliocentric velocities in km~s$^{-1}$ are indicated
in the top right corner of each panel. The ellipse indicates the size of
the synthesized beam. The contour levels are  -0.7, 0.7, 1.4, 2.1 mJy per 
beam. Negative contours are dashed.}
\end{figure}

\begin{figure}
\caption{ A position velocity profile along the major axis of 
F21569-330 at a position angle of -40$^{\circ}$.
Contour levels are -1,1,2,3 mJy per beam. Negative contours are dashed.} 
\end{figure}

\begin{figure}
\caption{ Contour images of the velocity channels of the galaxy north 
of PKS~2155-304. The cross indicates the optical position of the galaxy, 
heliocentric velocities (km~s$^{-1}$) are indicated in the top right corner
of each panel.  
Contour levels are -0.7, 0.7, 1.4 mJy  per beam. negative contours
are dashed.}
\end{figure}

\begin{figure}
\caption{ A position velocity profile along the major axis (east west)
of the galaxy north of PKS~2155-304.  
Contour levels are -0.7, 0.7, 1.4 mJy per beam. Negative contours
are dashed.}
\end{figure}

\begin{figure}
\caption{ An overlay of the H~I surface density distribution (contours)
of the three galaxies detected in H~I at velocities of about 17000 km~s$^{-1}$
on an optical image (greyscale), which shows the three galaxies and 
PKS~2155-304.
The contour interval is  2.34 $\times$ 10$^{19}$ cm$^{-2}$.   
The arrow points to PKS~2155-304.}
\end{figure}

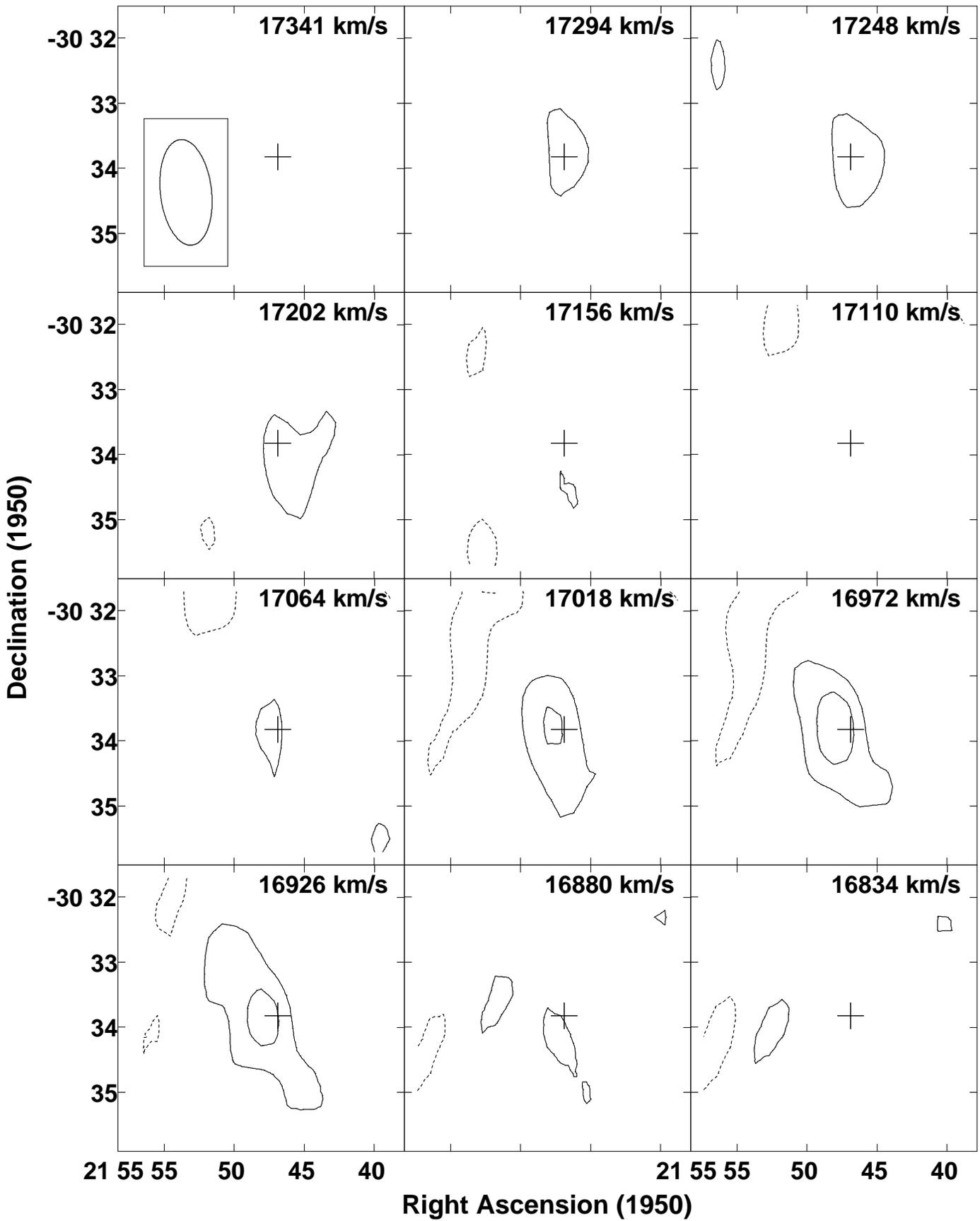
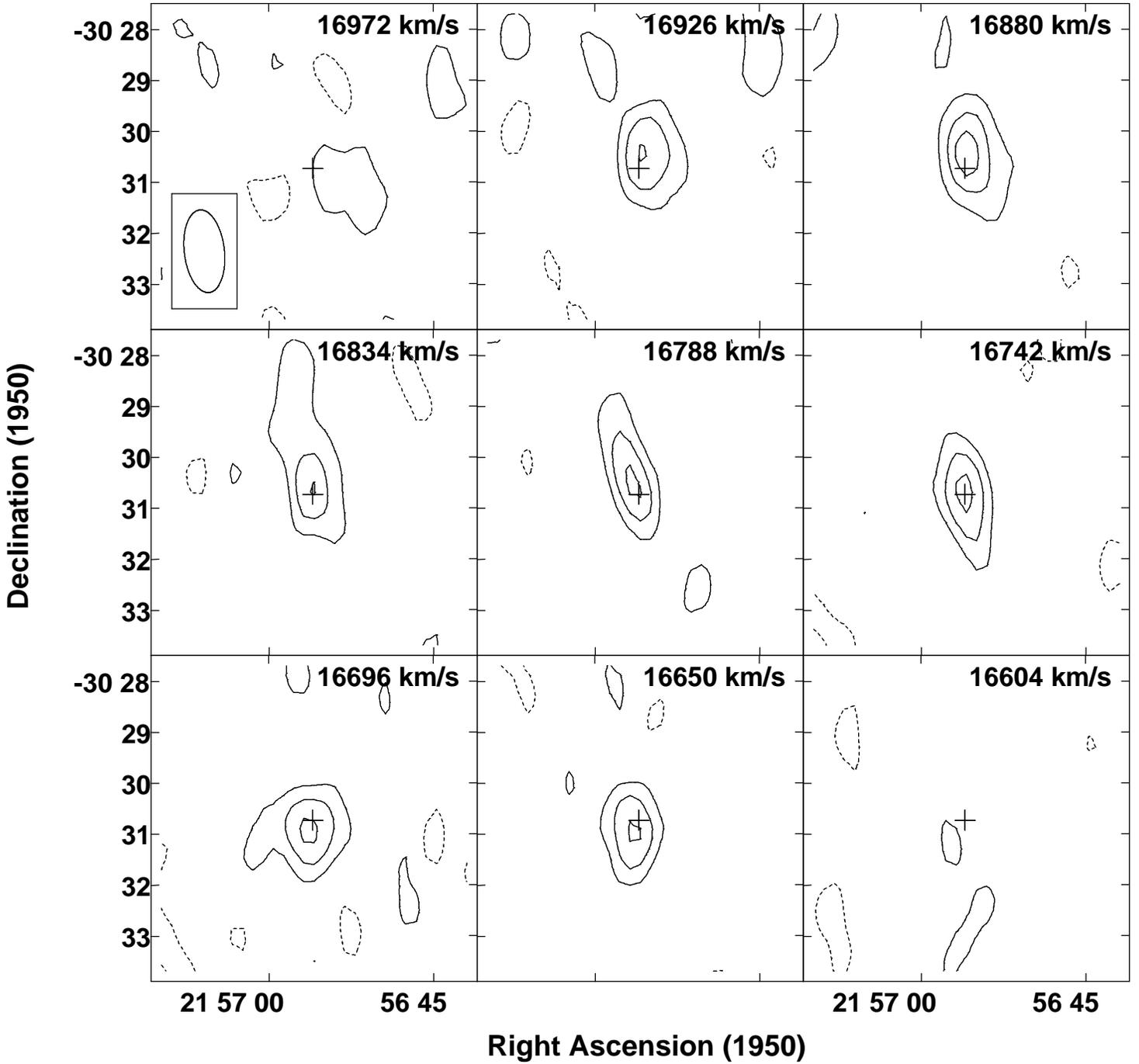
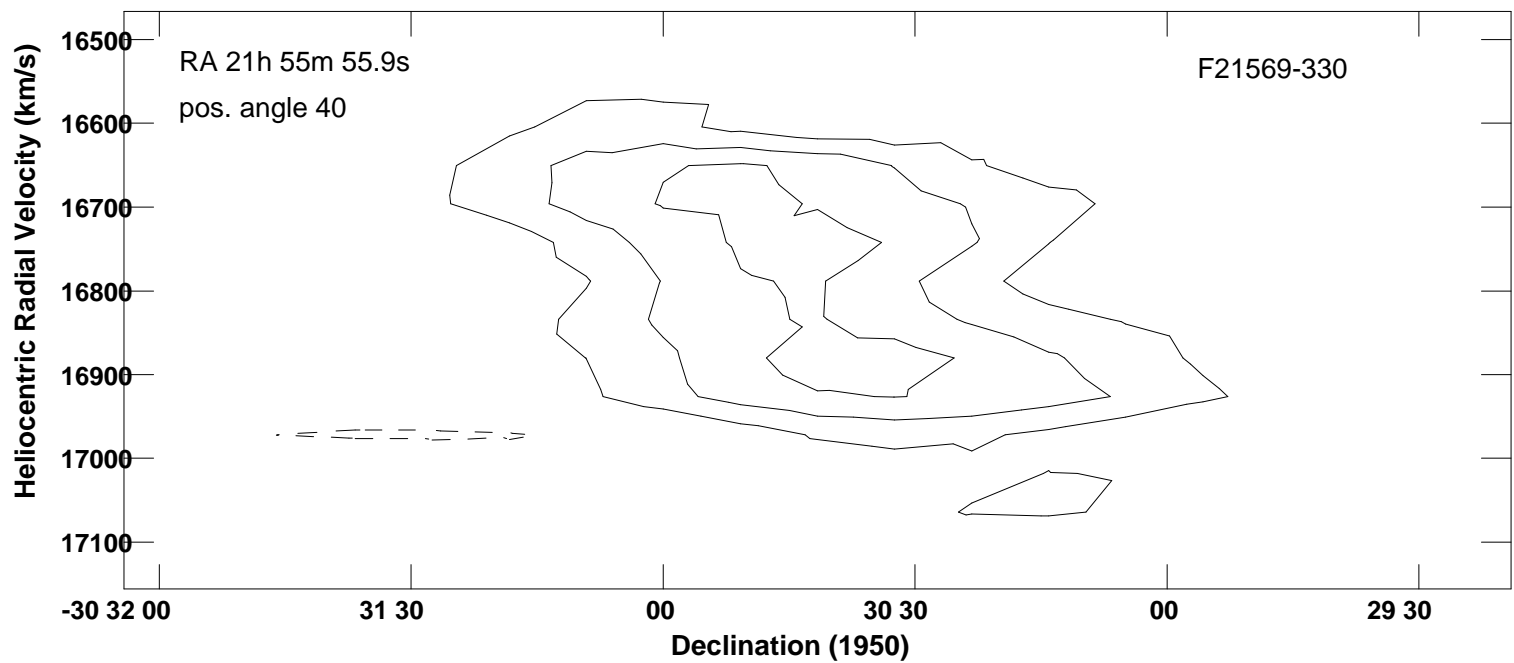
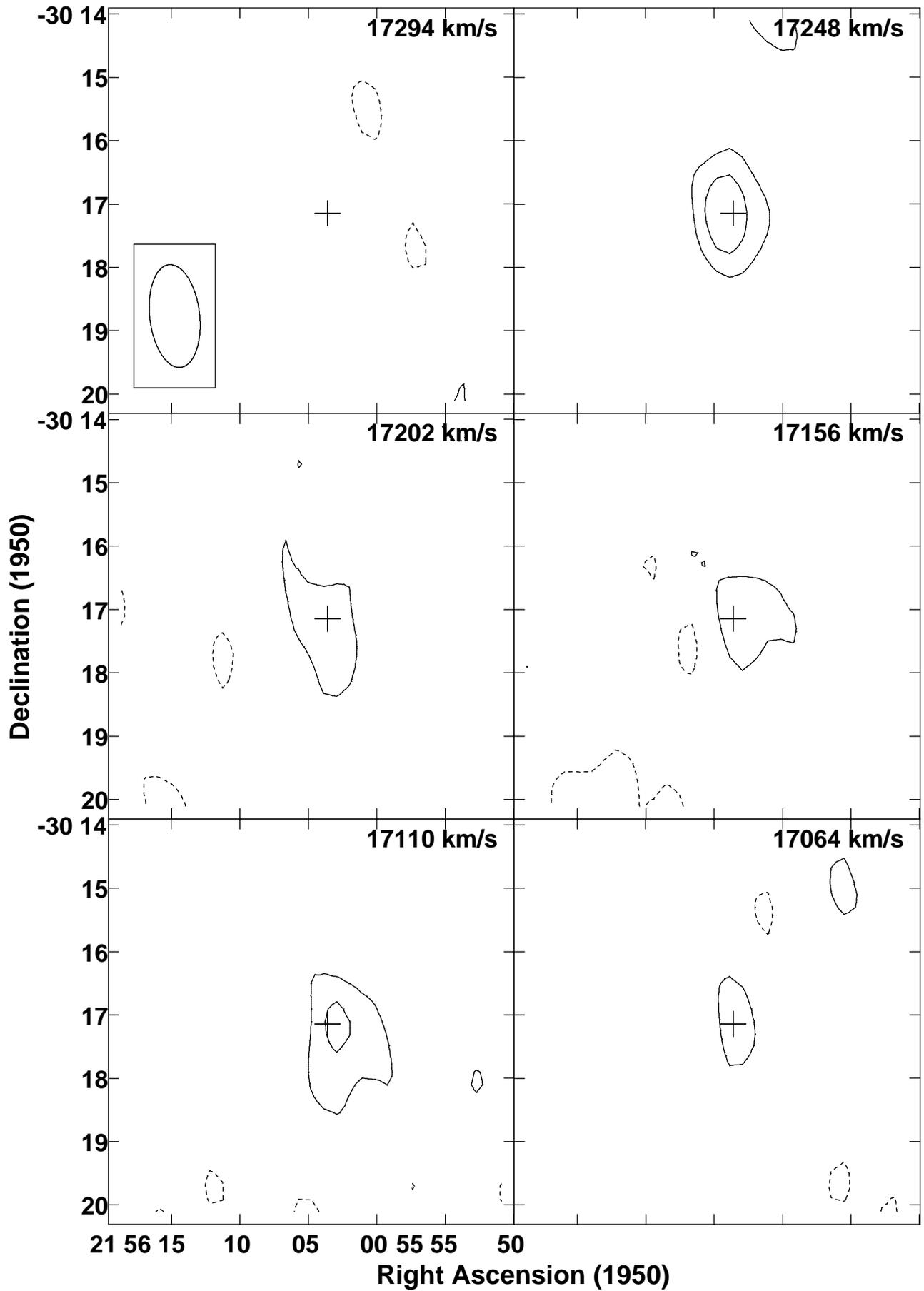
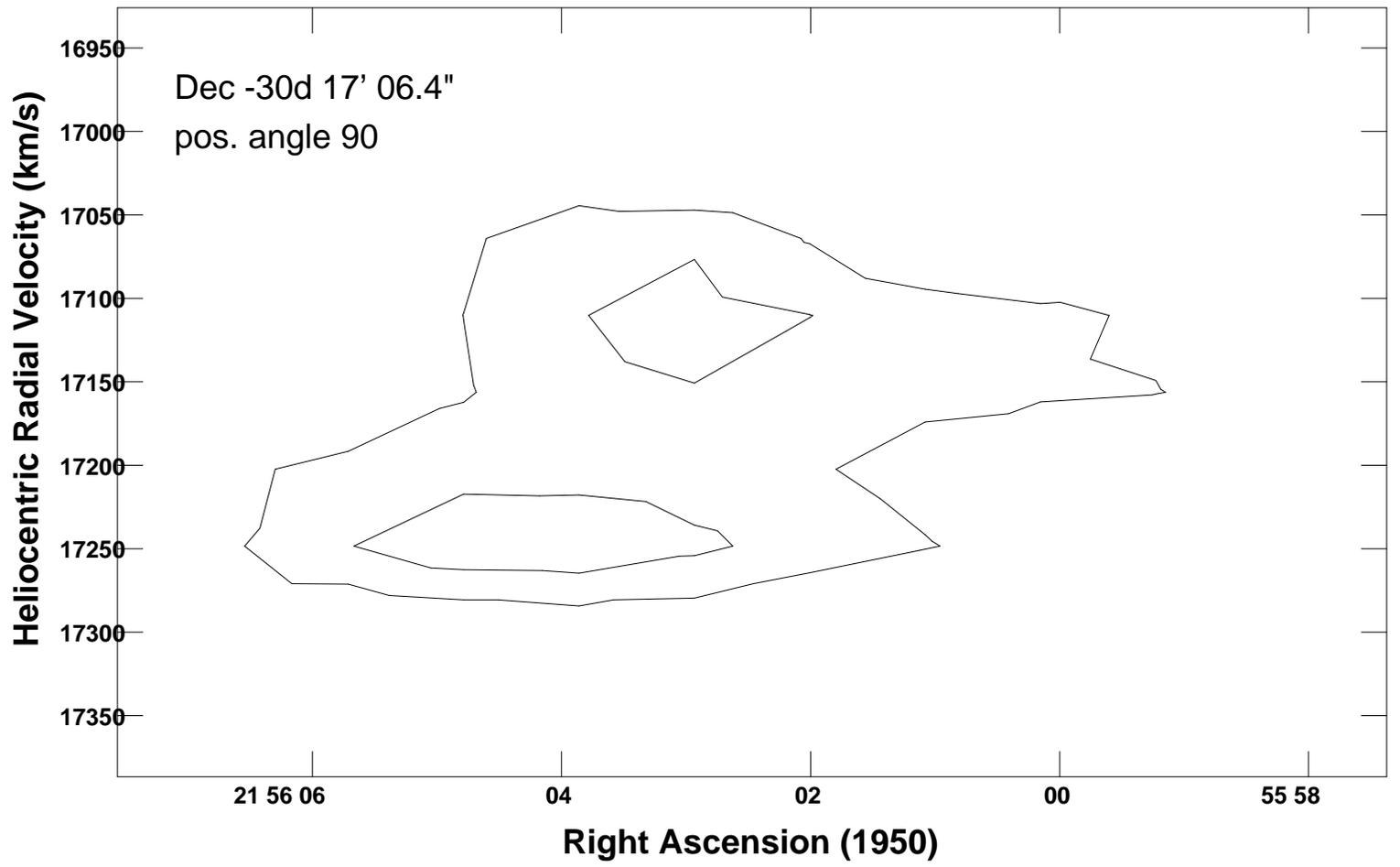
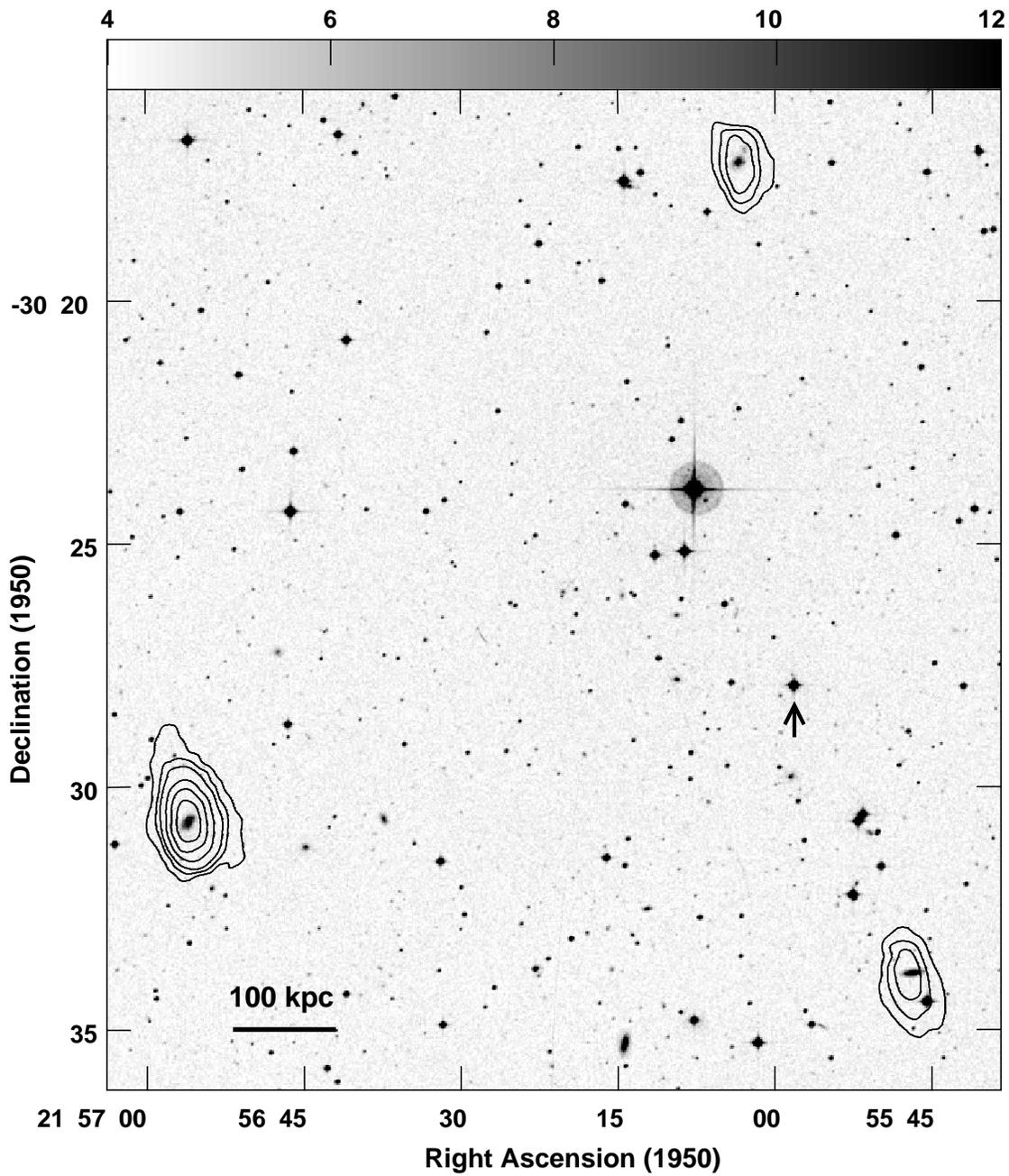
\end{document}